\documentclass[aps,prb,twocolumn,superscriptaddress,floatfix,goodfloat,showpacs]{revtex4}
\usepackage{graphics}
\usepackage{epsf}
\usepackage{bbm}

\newcommand \be {\begin{equation}}
\newcommand \ee {\end{equation}}
\newcommand \bea {\begin{eqnarray}}
\newcommand \eea {\end{eqnarray}}

\newcommand \bi {\bibitem}

\newcommand \lan {\langle}
\newcommand \ran {\rangle}

\newcommand{\eq}[1]{(\ref{#1})}

\newcommand{\myscalebox}[1]{\scalebox{0.35}[0.35]{#1}}

\begin{document}

\title{Order-Parameter Fluctuations in
Ising Spin Glasses at Low Temperatures}

\author{Matteo Palassini} \email{matteo@ipno.in2p3.fr}
\affiliation{Laboratoire de Physique Th\'eorique et Mod\`eles Statistiques,
B\^at. 100, Universit\'e  Paris-Sud, Orsay Cedex 91405, France}

\author{Marta Sales}
\affiliation{Department of Chemical Engineering, Northwestern University,
2145 Sheridan Rd., Evanston, IL-60202 (USA)}

\author{Felix Ritort}
\affiliation{Departament de F\'{\i}sica Fonamental, Facultat de F\'{\i}sica,
Universitat de Barcelona, Diagonal 647, 08028 Barcelona (Spain)}

\date{August 14, 2003}

\begin{abstract}
We present a numerical study of the order-parameter fluctuations for
Ising spin glasses in three and four dimensions at very low
temperatures and without an external field. Accurate measurements of
two previously introduced parameters, $A$ and $G$, show that the order
parameter is not self-averaging, consistent with a zero-temperature
thermal exponent value $\theta'\simeq 0$, and confirm the validity of the relation
$G=1/3$ in the thermodynamic limit in the whole low-temperature phase,
as predicted by stochastic stability arguments.
\end{abstract}

{\rm }
\maketitle

\section{Introduction}

Understanding the low-temperature physics of short-range spin
glasses~\cite{books} remains a major unsolved problem.
Much of the current debate 
concentrates on the equilibrium thermodynamics
of the Edwards-Anderson model with Ising spins (EAI model), the canonical
short-range spin glass. Since analytical approaches pose formidable
difficulties, the problem is often studied numerically.
However, the existence
of large barriers between low-energy configurations has limited so far
numerical calculations
to small systems sizes, from which it is hard to draw definite
conclusions on the large-volume limit.

Two main issues have been addressed in many numerical studies: the
existence and character of the finite-temperature spin-glass
transition, and the nature of the low-temperature spin-glass phase.  
A central quantity of interest in the description of the spin-glass
phase is the scaling exponent $\theta'$ governing the typical energy
of the lowest-lying excitations with linear size of order $l$, which
is assumed to scale as $E \sim l^{\theta'}$.  In general, $\theta'$
may be distinct \cite{KM00,PY00} from the stiffness exponent $\theta$
measured in domain-wall
computations~\cite{DW,HARTMANN,HARTMANN4D,PY99}, and the stability of
the spin-glass phase requires to have $\theta' \ge 0$.  In a
``many-state'' picture \cite{foot_states} such as the
replica-symmetry-breaking picture inspired by mean-field
theory~\cite{parisi,books,MPRRZ}, one has $\theta'=0$, hence there are
excitations whose energy remains finite (of the order of the coupling
strength between two spins) even as their length scale diverges.  In a
``two-state'' picture, such as the droplet model~\cite{fh,bm}, one has
$\theta' > 0$, hence the energy of large-scale excitations diverges
with their size. In this case the identity $\theta'=\theta$ is
often assumed.

Both the spin-glass transition and the ordered phase have been usually
investigated numerically by computing sample-averaged quantities such as the
Binder cumulant \cite{binder} or the distribution of the order
parameter (OP) and related observables.  
Recently~\cite{MNPPRZ,PPRi}, it was observed that
useful information on both issues can be drawn from the
sample-to-sample fluctuations of the OP.  In particular, two
dimensionless measures of the OP fluctuations were
considered~\cite{MNPPRZ,PPRi}: $A$, the normalized fluctuation of the
spin-glass susceptibility, and $G$, a ratio between two cumulants of
the OP distribution.  These two parameters are related to the Binder
cumulant, $B$, via the relation $B=1-A/(2 G)$. For a model without
time-reversal symmetry (TRS), $A, G$, and $B$ are given by
Eqs.~(\ref{eq2}), (\ref{eq3}), and (\ref{eq4}) below.  The parameter
$G$ serves as a good indicator of the existence of phase
transitions~\cite{PPRi} in systems lacking TRS (for which $B$ is
generally a bad indicator), as recently shown for several systems,
including Migdal-Kadanoff spin glasses~\cite{COMREP}, RNA folding
models~\cite{PPRICCI}, chiral spin systems~\cite{HK}, and mean-field
models such as the SK model (with and without a magnetic field), the
infinite-range $p$-spin model~\cite{PRS}, and the infinite-range Potts
model~\cite{POTTS}.  The parameter $A$ has also been studied before
for random diluted models at criticality~\cite{WD96,AH}.

In this paper, we investigate the OP fluctuations in the EAI
model with Gaussian couplings in three and four dimensions by
low-temperature Monte Carlo simulations.  We study the case with no
external field, which satisfies TRS.
In three dimensions (3D), numerical data are available
in the literature for $A$ in the high-temperature phase~\cite{MPR94} and
for $G$ near the critical point~\cite{BCF}, for the ``$\pm J$'' 
coupling distribution.
In four dimensions (4D), $G$ was measured  at moderately low
temperatures, also for the $\pm J$ distribution~\cite{MNPPRZ}.
Here, we study much
lower temperatures than in these studies, in order to reduce
crossover effects associated to the critical point~\cite{BDM,KPY},
which complicate the interpretation of the numerical data at
higher temperatures.

A summary of our results is as follows.  First, we
estimate $\theta'$ from the system-size dependence of $A$, finding,
for the system sizes we could reach, a small value of $\theta'$
incompatible with the accepted values of the domain-wall exponent
($\theta \simeq 0.2$ in
3D~\cite{DW,HARTMANN,PY99} and $\theta \simeq0.7$ in
4D~\cite{HARTMANN4D}), and compatible with zero.
  This agrees with recent determinations of $\theta'$ from
ground-state perturbation methods~\cite{KM00,PY00,MP01} and from
low-temperature measurements of the OP
distribution~\cite{KPY,RB90,MZ,MPRRZ} (which all consider
sample-averaged quantities), and supports a picture of the spin-glass
phase characterized by two distinct exponents, $\theta >0$ and
$\theta'=0$. The result $\theta'=0$
implies that the OP is not {\em self-averaging} in the thermodynamic limit.

Second, we find good evidence that the
identity $G=1/3$ holds in the whole spin-glass phase in the
thermodynamic limit, confirming the validity of sum rules proposed
by Guerra~\cite{GUERRA1} and first derived for the SK model, which follow
from the property of ``replica equivalence''~\cite{BMY,MPSV}.

Third, we find that $A$ and $G$ allow to locate the spin-glass
transition reasonably well although, as expected due to TRS and as
previously numerically observed~\cite{BCF}, $B$ provides a better
determination (a much more accurate determination is provided
by the correlation length~\cite{BCF}, which we do not
investigate here).

We do not study in this paper the surface fractal
dimension $d_s$ of the excitations, which is the other exponent,
besides $\theta'$, characterizing the spin-glass phase (in 
particular, $d_s=0$ in the standard
replica-symmetry-breaking picture \cite{MPRRZ}, $d_s>0$ in the droplet 
model, while the ``TNT'' picture \cite{KM00,PY00} predicts the
``mixed'' behavior $d_s>0$, $\theta'=0$).

The rest of the paper is organized as follows. In Sec.~II we introduce the 
different
models and observables studied, and discuss the theoretical predictions
for these observables. In Sec.~III we present and analyze
our numerical results for the quantities $A$, $G$ and $B$. Finally,
in Sec.~IV we summarize our conclusions.

\section{Models, observables, and theoretical predictions}

We study the EAI model defined by the Hamiltonian
\be {\cal H}=-\sum_{\langle i,j\rangle} J_{ij}
S_i S_j\,,~~~S_i=\pm1~~,
\label{eq1}
\ee
where $L^D$ Ising spins $S_i$ sit on a (hyper-)cubic lattice in $D$
dimensions with linear size $L$ and periodic boundary conditions in all 
directions.
The couplings $J_{ij}$ are drawn
from a Gaussian distribution of zero mean and unit variance. We
consider two different models:
({\em i}) the case with interactions $\langle i,j \rangle$ restricted to nearest neighbors
(referred to as NN model) in $D=3$ and $4$; 
({\em ii}) the case with interactions
restricted to nearest, next-nearest, and next-next-nearest neighbors 
(referred to as NNN model) in $D=3$,
which has a coordination number $z=26$.

The NN model has been extensively studied and is known to display a
finite-temperature continuous spin-glass transition for $D\geq 3$. Recent estimates
of the critical temperature for Gaussian-distributed couplings are $T_c=0.95\pm 0.04$ in
3D~\cite{MPR} and $T_c=1.80\pm 0.03$ in 4D~\cite{PRR96}.

The NNN model has been much less studied.
In Ref.~\onlinecite{MPR94}, the 3D case with $\pm J$ couplings
was considered, but
no conclusive evidence of a finite-temperature transition
was obtained, the data being compatible with both
$T_c\approx 3.27$ and a zero-temperature
singularity. Incidentally, in the NNN case we do not expect a large difference
in $T_c$ between binary and Gaussian distributions, due
to the large coordination number.
For the NNN model, we did not consider temperatures as
low as for the NN model, but focused on the phase
transition region. Another of the results of this paper is
a convincing evidence that indeed a finite-temperature
transition exists in this model in 3D.

We measure $A$, $G$, and $B$ as a function of temperature $T$ and
size $L$ using the following definitions:
\begin{eqnarray}
A(T,L)&=&\frac{\overline{\langle q^2\rangle^2}-\overline{\langle
q^2\rangle}^2}
{\overline{\langle q^2\rangle}^2} \;
\label{eq2}
\\
G(T,L)&=&\frac{\overline{\langle q^2\rangle^2}-\overline{\langle
q^2\rangle}^2}
{\overline{\langle q^4\rangle}-\overline{\langle q^2\rangle}^2} \;
\label{eq3}
\\
B(T,L)&=&\frac{1}{2}\left(3-\frac{\overline{\langle
q^4\rangle}}{\overline{\langle q^2\rangle}^2}\right) \;  \label{eq4}
\eea
where
$q=\frac{1}{L^D} \sum_{i} S_i^a S_i^b $ is the spin overlap of two
independent systems $S_i^a$ and $S_i^b$ with the same random
couplings, and $\lan ...\ran$ and $\overline{(...)}$ stand for
thermal and disorder averages, respectively. The OP for
a given realization of the
disorder is $\lan q^2 \ran$, therefore $A$ is nothing but
the normalized sample-to-sample variance of the OP.

In the paramagnetic phase, $T > T_c$, and for sufficiently large $L$
so that $L\gg \xi$ (where $\xi$ is the correlation length),
the OP follows a Gaussian distribution and all three parameters 
vanish as $1/L^D$, as follows from the central limit
theorem. Following the terminology of Wiseman and Domany \cite{WD96},
this means that the OP is {\em strongly self-averaging}.

At $T=T_c$, the correlation length diverges and the central limit theorem
cannot be applied. 
For strongly disordered systems such as spin glasses it is known that
the OP is not self-averaging at criticality \cite{WD96,AH}, 
namely $A$ tends to a finite value in the thermodynamic limit. 
If $A$ is finite then clearly  $G$ must
be finite, and standard renormalization-group arguments show that $B$ 
is also finite at $T_c$. Since $B$ and $G$ are dimensionless and monotonic
in $T$, in plots of these quantities as a function of $T$, the curves
for different values of $L$ must all
cross at $T=T_c$, and one can use this
to determine $T_c$.  From standard finite-size scaling, one can then
determine the critical exponent $\nu$. Since much work has been
devoted to measuring $\nu$ from the standard observables (see for
instance Refs.~\onlinecite{YOUNG,BCF,PC,ky,MPR}), and we are primarily 
interested
in the low-temperature phase here, we will not attempt a precise
determination.

In the spin-glass phase, $T<T_c$, $A$ is expected~\cite{RSb,DBM} to
vanish linearly with $T$ according to the scaling law
\be A(T,L) \sim T\, L^{-\theta'}~~,
\label{asc}
\ee
where $\theta'$ is the exponent discussed in the Introduction. This law
holds under two hypothesis (both satisfied in the case of continuous
couplings studied here): ({\em i}\/) the ground state is unique; ({\em
ii}\/) the
probability distribution of the energy  of the 
lowest-lying excitations has finite weight at
zero energy~\cite{RSb,DBM}.

From the above scaling law we see that if $\theta'>0$, then $A$
vanishes for $L\to \infty$, namely the OP is {\em weakly
self-averaging} (where ``weakly'' indicates \cite{WD96} 
that the OP fluctuations
vanish more slowly than $1/L^d$, a consequence of  the inequality
$\theta' < d$). This situation is encountered in the droplet
model~\cite{fh,bm}, as discussed in the Introduction, and also
in mean-field models with a marginally stable replica-symmetric solution
at low temperatures (such as the spherical
SK model~\cite{RS}). If $\theta'=0$, as in a ``many-state'' picture,
$A$ remains finite in the thermodynamic limit, namely the OP is
{\em not} self-averaging.

Turning now to $G$, it is known \cite{BMY,MPSV} 
that in the SK model the following
relation holds for $T < T_c$:
\begin{equation} 
\lim_{L\to \infty} G(T,L)=1/3  \, .
\label{guerra}
\end{equation}
According to Guerra~\cite{GUERRA1}, this relation should hold (for $T<T_c$)
in any model which is ``stochastically stable'' with respect to a
mean-field perturbation and which has a non-self-averaging OP.  Under
the hypothesis ({\em i}\/) and ({\em ii}\/) above, the more general
conjecture has also been made~\cite{RS} that the above relation holds
for $T<T_c$ even if the OP is self-averaging. In this case,
$G$ would be finite but both the numerator and the denominator in
Eq.(\ref{eq3}) would vanish, as for example in the
Migdal-Kadanoff spin glass (see Bokil et al. in
Ref.~\onlinecite{COMREP}) and the SK spherical model~\cite{RS}. It has 
also been explicitly proven~\cite{RSb}, under the hypothesis ({\em i}\/) 
and ({\em ii}\/),  that one has $G(T=0,L)=1/3$ for any $L$. Note that models in which
$G(T=0,L)\ne 1/3$ in general will not satisfy the 
conjecture of Ref.~\onlinecite{RS}.

\section{Numerical results}

\begin{table}
\begin{center}
\begin{tabular*}{\columnwidth}{@{\extracolsep{\fill}}r|llllll}
\hline
\hline
Model & $L$ &  $T_{\mbox{min}}$ &  $T_{\mbox{max}}$ & $N_T$ & $N_s$ & MCS \\
\hline
3D NN  & 4  & 0.1  & 2.0 & 18 & 16000& $10^5$\\
       & 6  & 0.2  & 2.0 & 16 & 6000& $10^5$\\
       & 8  & 0.2  & 2.0 & 16 & 6600& $10^5$\\
       & 12 & 0.94 & 2.0 & 14 & 3751& $3 \cdot 10^5$\\
       & 16 & 0.94 & 2.0 & 16 & 587& $10^6$\\
\hline
4D NN  & 3  & 0.2  & 2.8 & 12 &16000 & $10^4$\\
       & 4  & 0.2  & 2.8 & 12 &13951 & $10^5$\\
       & 5  & 0.46  & 2.8 & 19 & 1476& $3 \cdot 10^5$\\
       & 7 & 0.995 & 2.8 & 29 & 832& $3 \cdot 10^5$\\
\hline
3D NNN  & 4  & 2.0  & 5.0 & 16 & 9005& $10^4$\\
       & 6 & 2.0  & 5.0 & 16 & 3258& $10^4$\\
       & 8  & 2.0 & 5.0 & 16 & 3574& $3 \cdot 10^4$\\
       & 12 & 2.8 & 5.0 & 12 &1751& $10^5$\\
       & 16 & 3.4 & 5.0 & 9 &489&  $10^5$\\
\hline
\hline
\end{tabular*}
\end{center}
\caption{Parameters of the simulations. $L$ is the linear system size,
$T_{\mbox {min}}$ and $T_{\mbox {max}}$ the smallest and largest 
temperatures considered, $N_T$ the number of temperatures in the 
parallel tempering algorithm, $N_s$ the number of independent 
realizations of the disorder (samples), and MCS the number of
Monte Carlo steps per spin and per temperature.}
\label{tab:param}
\end{table}

We simulated the various models with the
parallel tempering technique~\cite{HN}, which allows to reach significantly
lower temperatures than conventional Monte Carlo methods.
The parameters of the simulation are given in Table~\ref{tab:param}.
Equilibration of the Monte Carlo runs was tested by monitoring 
all the measured observables on a logarithmic time scale, checking
that they had 
all converged within their statistical errors, and
by applying the equilibration test discussed in Ref.~\onlinecite{KPY}.

\subsection{Parameter $A$}

In Figures~\ref{A3NN}, \ref{A4NN}, and \ref{A3NNN} we show our numerical
results for $A$ in the 3D NN, 4D NN, and 3D NNN models, respectively.
The vertical lines in Figures~\ref{A3NN} and \ref{A4NN} 
indicate the estimated value of $T_c$.
In all cases, the behavior of $A$ resembles that observed in the SK
model (see Refs.~\onlinecite{MNPPRZ,HK,PRS}).  At high temperatures, 
$A$ decreases with
$L$, approximately as $1/L^D$, showing that the OP is 
strongly self-averaging in this regime,  as expected.  Near $T=T_c$, 
there is a maximum whose
position shifts towards $T_c$ as $L$ increases, an effect of
finite-size corrections.  The shift is modest in the 4D NN model but
 quite noticeable in the 3D NN model, where even  for the
largest $L$ the position of the maximum is still
significantly larger than $T_c$. In the 3D NNN model, the position of
the maximum is also larger than $T_c$ (see discussion in 
Sec.~\ref{sub:critical} on the value of $T_c$ in this model), 
but the shift is
less pronounced.  The {\em height} of the maximum increases with $L$ in all
models, indicating that $A$ attains a finite value in 
the thermodynamic limit (since it is
bounded from above), namely that the OP is not self-averaging at $T_c$, as
expected.

At low temperatures ($T<T_c$), Figures~\ref{A3NN}, \ref{A4NN}, and \ref{A3NNN} 
show that $A$
is approximately linear  in $T$, in agreement with Eq.~\eq{asc}.
Most interestingly, the data for different values of $L$ tend to
superimpose to each other. In a scenario with $\theta'>0$, 
the data should tend to zero for large $L$ in the whole region below $T_c$.
In 3D, we see no decrease at all in the data with increasing $L$,
while a modest decrease is observed in 4D.
\begin{figure}
\myscalebox{\includegraphics{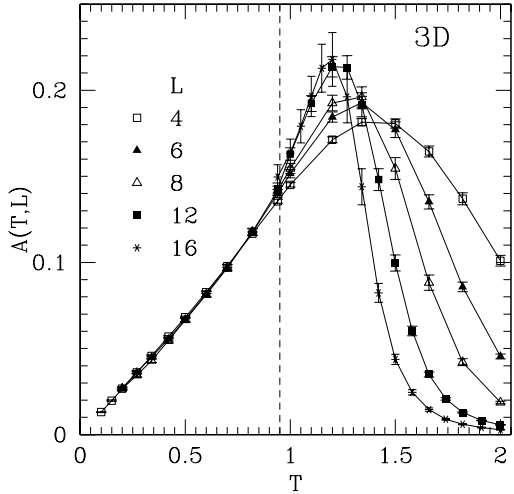}}
\caption{Parameter $A$ for the 3D NN model as a function of the temperature,
for different system sizes $L$. The vertical line represents the
estimated value of the critical temperature, $T_c=0.95\pm 0.04$.}
\label{A3NN}
\end{figure}
\begin{figure}[h!]
\myscalebox{\includegraphics{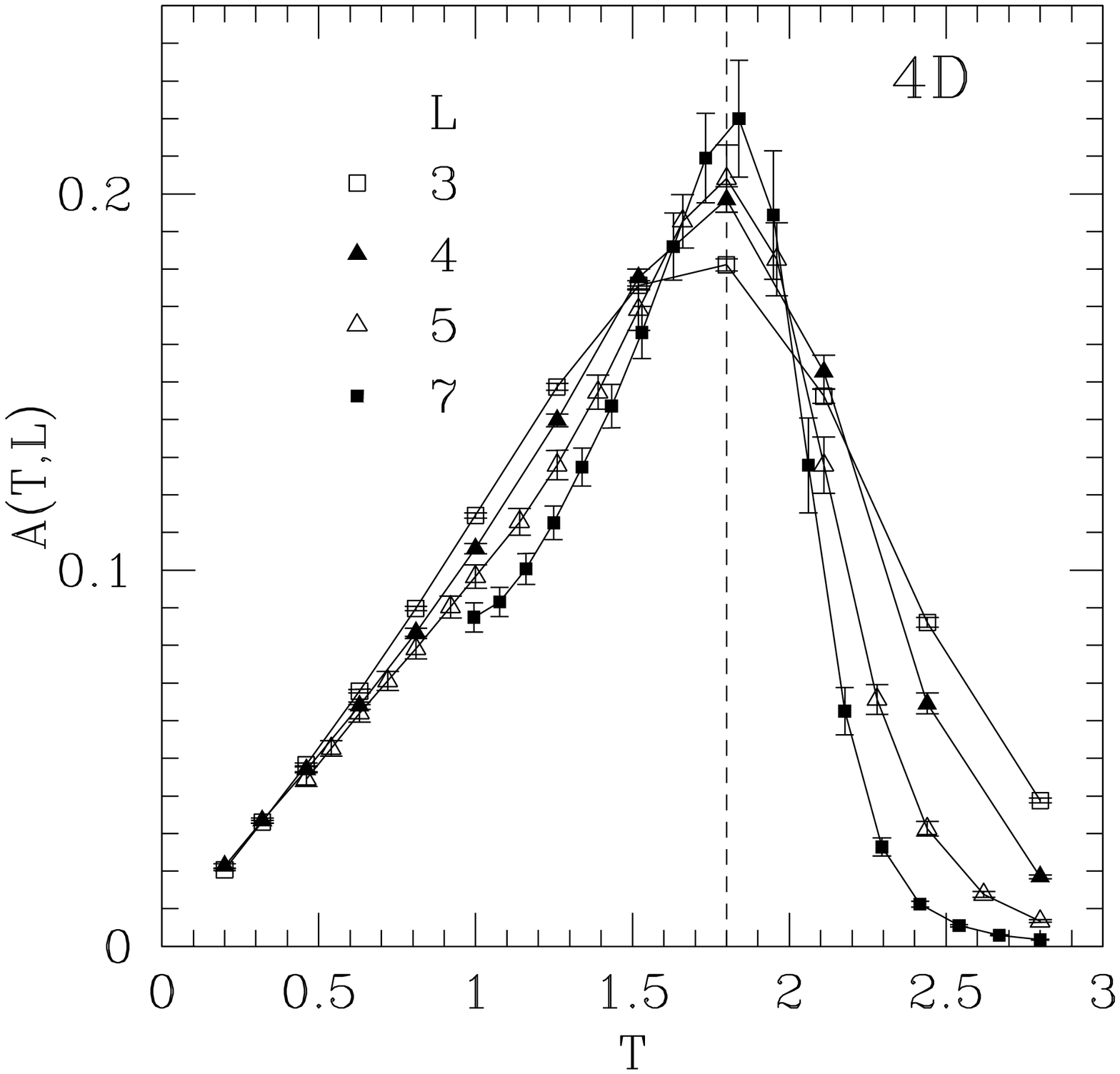}}
\caption{Same as Figure~\ref{A3NN} but for the 4D NN model. The vertical
line corresponds to $T_c=1.80\pm 0.03$. \label{A4NN}}
\end{figure}
\begin{figure}[h!]
\myscalebox{\includegraphics{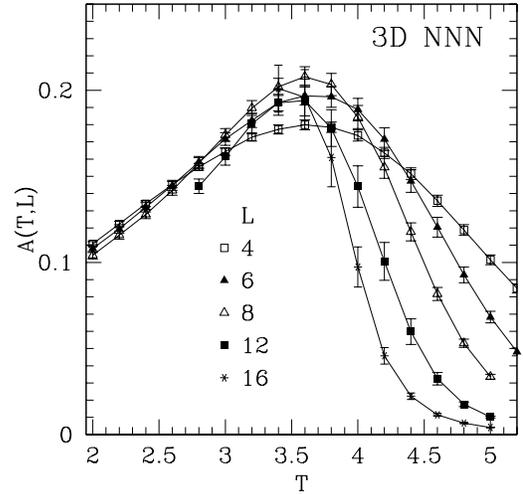}}
\caption{Same as Figure~\ref{A3NN} but for the 3D NNN model.}
\label{A3NNN}
\end{figure}

To analyze in more detail the size dependence of $A$ at low temperatures,
in Figure~\ref{A34L} we
plot the ratio $A/T$ as function of $L$ at different temperatures for the
three models. The straight
lines represent the scaling law Eq.(\ref{asc}) assuming 
$\theta'=\theta$,
and using the estimates of $\theta$ from domain-wall
calculations, $\theta = 0.2$ in
3D~\cite{DW,HARTMANN,PY99} and $\theta = 0.7$ in
4D~\cite{HARTMANN4D}. No
estimates of $\theta$ are available for the 3D NNN model, so we use that
for the NN model  (we expect that $\theta$ is a universal
exponent equal for both models).
Clearly, the data in Figure~\ref{A34L} do not agree with the hypothesis 
$\theta'=\theta$
for the range of sizes considered, and seem to saturate to a constant
value instead. We fitted the data with the form $A(T,L)/T =
a L^{-\hat{\theta}'}$, where $\hat{\theta}'$ should be seen 
as an ``effective'' exponent which depends on the temperature 
and which  effectively takes into account
corrections to the leading scaling behavior, 
with $\hat{\theta'}\to \theta'$ in the limit
$T L^{-\theta'} \to 0$.
The fits give $\hat{\theta'}$ varying from
$0.03 \pm 0.02$ ($T=0.7$) to  $0.00\pm 0.06$ ($T=0.2$) in the 3D NN model,
from $0.30\pm 0.05$ ($T=1.0$) to  $0.003\pm 0.006$ ($T=0.32$) in the 
4D NN model, and from $0.03\pm 0.04$ ($T=2.8$) to  $0.08\pm 0.04$ ($T=2.0$) 
in the 3D NNN model.

Therefore, in all cases the data is
compatible with $\theta'= 0$, in agreement with the ``many-state''
picture, and is statistically incompatible with with $\theta'=\theta$,
in disagreement with the ``two-state'' picture.
As usual, we cannot exclude
a crossover~\cite{BDM} to a larger value of $\theta'$
for larger $L$. In this case, in the large-volume limit $A$ would be
zero at all temperatures, except at $T=T_c$.

A value of $\theta'$ compatible with zero was also obtained from the OP
distribution~\cite{KPY,RB90,MZ,YOUNG,MPRRZ}
and from direct measurements of the energy of low-lying excitations
created by perturbing the ground state~\cite{KM00,PY00,MP01}.

\begin{figure}[h!]
\begin{center}
\myscalebox{\includegraphics{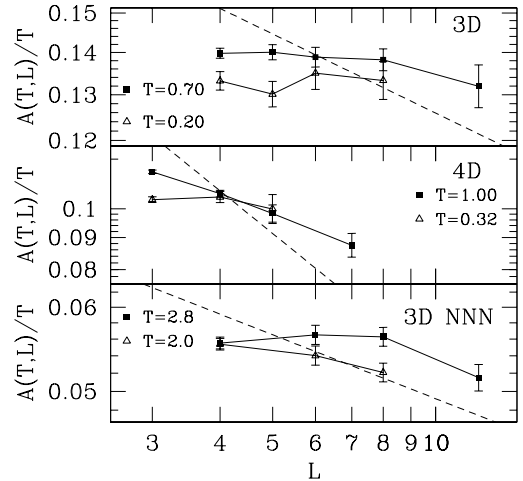}}
\end{center}
\caption{Log-log plot of $A/T$ versus $L$
at different temperatures for the 3D NN model
(top), the 4D NN model (middle), and the 3D NNN model (bottom). 
The straight lines of slope -0.2 (top and bottom)
and -0.7 (middle) are the expected
scaling behavior of Eq.(\ref{asc}) if $\theta'=\theta$.~\label{A34L}}
\end{figure}

\subsection{Parameters $G$ and $B$}

Figures~\ref{bg3nn}, \ref{bg4nn} and \ref{bg3nnn} show our
numerical results for $G$ and $B$ for the three models.  At high 
temperatures,
$G$ and $B$ vanish approximately as $1/L^D$ in all three models, again indicating
strong self-averaging.
Near  $T=0$, the data
for the NN model in both 3D and 4D
are compatible with $G(T=0,L)=1/3$ and $B(T=0,L)=1$, as
expected~\cite{RS,RSb} for a continuous coupling distribution.
More importantly, $G(T,L)$ seems to
converge to the value 1/3 for large $L$ in the {\em whole}\/
low-temperature region, in agreement with Eq.(\ref{guerra}).
This is particularly evident in the
4D NN model (Figure \ref{bg4nn}), where $G(T,L)$ has already converged to
1/3 for $L=5$ at
temperatures below $T \approx T_c/2 \approx 0.9$
(the data points above $1/3$ are due to
incomplete equilibration~\cite{PRS}). It is quite unlikely that the
saturation is a finite-size effect, therefore our results strongly
suggest that indeed $G=1/3$ in the whole spin-glass
phase. As discussed above, if $A$ remains finite as $L\to\infty$
(as indicated by our data),
this is an expected consequence of  stochastic stability
~\cite{GUERRA1}. If $A$ vanishes, instead,
our results would support the more general conjecture
of Ref.~\onlinecite{RS}.

\subsection{Critical region}
\label{sub:critical}

In this subsection we comment on the behavior of $G$ and $B$ near the
critical temperature, starting from the NN model.

In 3D, the vertical lines in Figure~\ref{bg3nn} indicates the position
of the critical temperature, using the value $T_c=0.95\pm 0.04$
quoted in Ref.~\onlinecite{MPR}, which was obtained from the
parameter $B$ measured in a large-scale simulation. One sees
that the data for both $B$ and $G$ for different values of $L$ come together
as $T$ approaches $T_c$ from above, as indicative of a phase
transition. Below $T_c$, the data for $G$ separate again in
a statistically significant way, while for $B$
one would need a substantially larger statistics (or larger sizes)
to see a clear separation, as observed in previous studies \cite{MPR,ky}.
For example, at $T=0.82$ the separation between  the $L=12$ and
the $L=4$ data is 1.4 standard deviations for $B$ and 2.5 standard
deviations for $G$.  The small separation below $T_c$ is probably
due to the vicinity of $D=3$ to the lower critical dimension
\cite{MPR,ky,PC}.
We also note that the crossing point of $G$ is at
temperatures larger than $T_c$, and close inspection shows that
it shifts towards $T_c$ from above as $L$ increases.
A similar shift was observed for the position of the
maximum of $A$ in Figure~\ref{A3NN}.

In 4D, both $B$ and $G$ display a very clear crossing (see Figure 
~\ref{bg4nn}),
as also observed in previous studies \cite{PRR96,MZ}. From $B$
we estimate $T_c=1.80 \pm 0.03$ in agreement with the results of 
Refs.~\onlinecite{BY,PRR96}. This value is indicated by the vertical lines
in Figure~\ref{bg4nn}. As in 3D, the crossing point of $G$
is at temperatures larger than $T_c$ and shifts towards $T_c$
as $L$ increases. 

Overall, this confirms that both in 3D and 4D the
corrections to scaling are significantly larger for $G$ and $A$ than for 
$B$.
Since $G$ and $A$ have also much larger statistical errors than $B$,
the latter quantity is to be preferred to $G$ and $A$ for
locating $T_c$ in models with TRS. As already mentioned, a much more
accurate quantity for this purpose is the correlation length, which shows
a very clear crossing in 3D\cite{BCF}, unlike $B$ and $G$.

Finally, in the 3D NNN model both $G$ and $B$
show a rather clear crossing (see Figure \ref{bg3nnn}).
This provides a clear evidence for the existence of a phase transition
in 3D Ising spin glasses, confirming recent results for
the NN model \cite{ky,PC,BCF} that
obtained a convincing evidence (especially Ref.\onlinecite{BCF}) after
the issue had remained unsolved for a long time.
The crossing point is at $T \simeq 3.3$ for $B$ and at somewhat higher
temperatures for $G$, although also here the crossing for $G$ shifts
to the left as $L$ increases.
From the data for $B$ one might  be tempted to conclude that the critical
temperature is $T_c\simeq 3.3$.
However, if this was the case, the value of $B$ at $T_c$ (which is a 
universal quantity)
would be lower in the 3D NNN model than in the 3D NN model, 
violating
universality. This suggests that the actual value of $T_c$ is significantly 
lower
than 3.3, despite the clear crossing of $B$
(which would then be strongly affected by scaling corrections), and for this reason
we have not indicated the position of $T_c$ in Figures \ref{A3NNN} 
and \ref{bg3nnn}.
A more detailed analysis \cite{matteo_preparation}
clearly shows that indeed $T_c$ is significantly lower than 3.3
in this model.

\begin{figure}[h!]
\myscalebox{\includegraphics{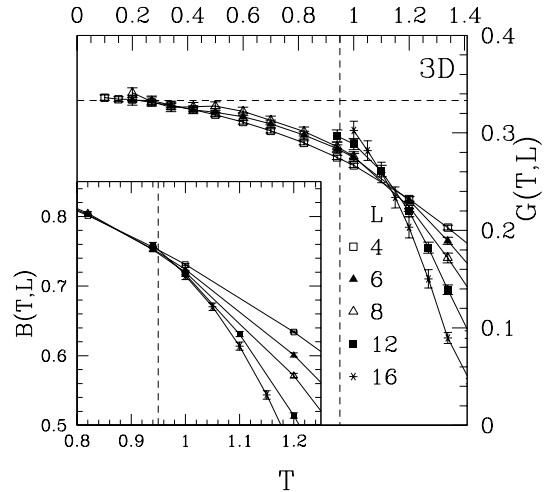}}
\caption{Parameters $G$ (main figure) and $B$ (inset) 
for the 3D NN model, as a function of the temperature and for various 
system sizes. The vertical lines
correspond to $T_c=0.95 \pm 0.04$, the horizontal line in the main figure
corresponds to the limit relation $G=1/3$. }
\label{bg3nn}
\end{figure}
\begin{figure}[h!]
\myscalebox{\includegraphics{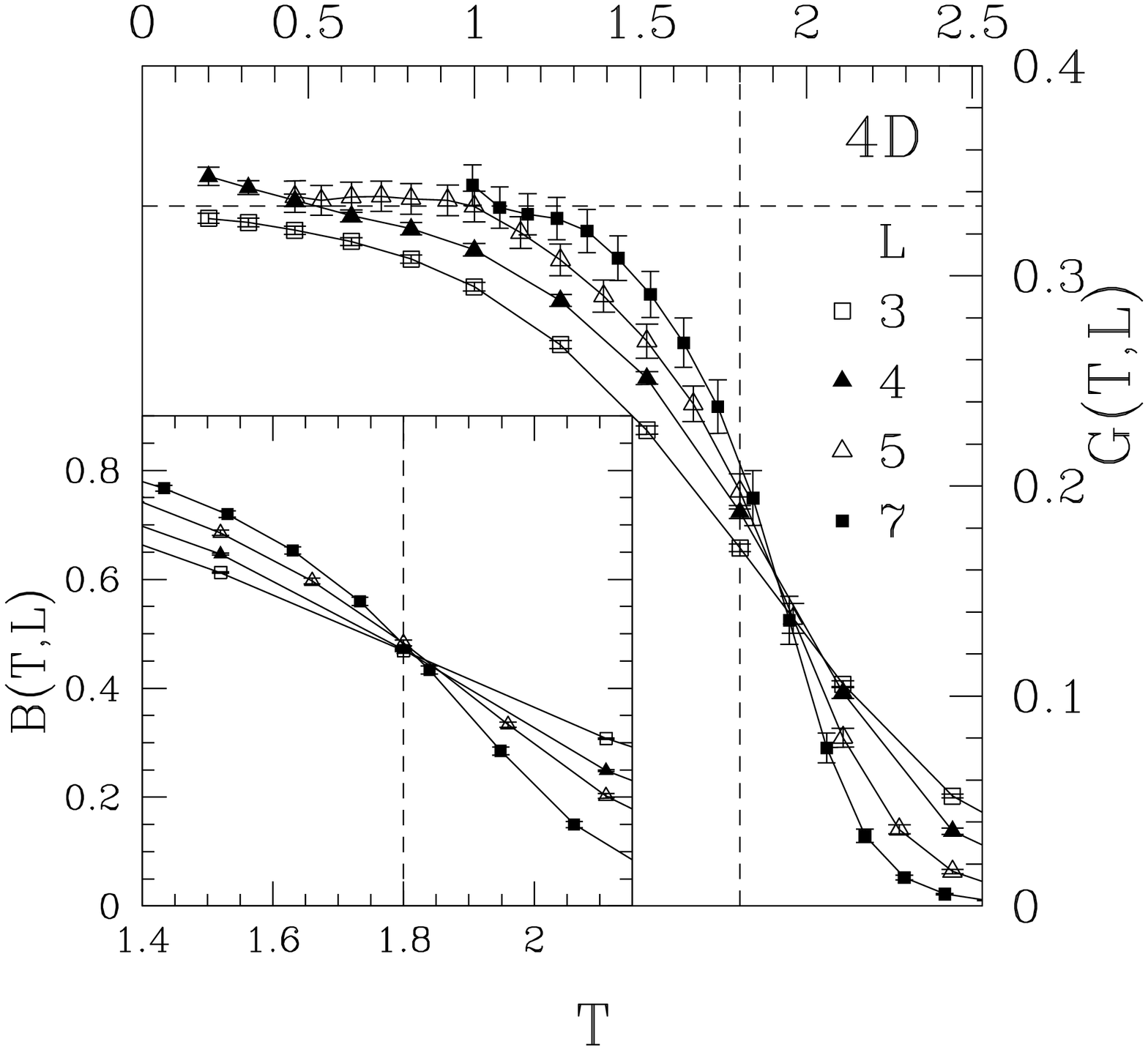}}
\caption{Same as Figure~\ref{bg3nn} but for the 4D NN model.
The vertical lines correspond to $T_c=1.80\pm 0.03$.}
\label{bg4nn}
\end{figure}
\begin{figure}[h!]
\myscalebox{\includegraphics{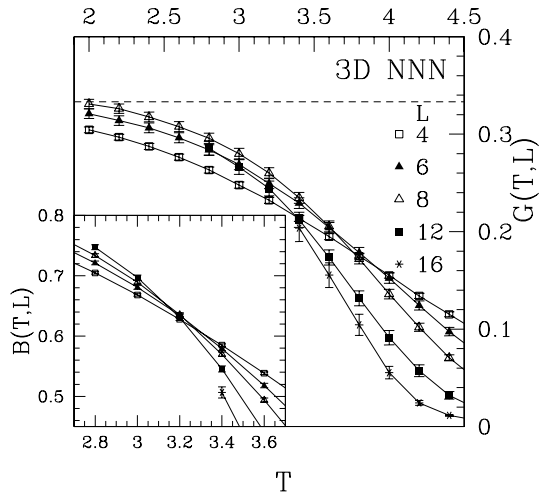}}
\caption{Same as Figure~\ref{bg3nn} but for the 3D NNN model. The true critical
temperature is significantly lower than the crossing point of $B$.}
\label{bg3nnn}
\end{figure}

\section{Conclusions}
To conclude, we have provided evidence that the order parameter is not
self-averaging in the low-temperature phase of the
Edwards-Anderson Ising spin glass in 3D and 4D,
which implies an exponent $\theta'\simeq 0$, in agreement with a
``many-state'' picture of the spin-glass phase, such as the
replica-symmetry-breaking picture or the ``TNT'' picture.  
As usual, due to the
limited system sizes that are currently reachable in numerical simulations,
we cannot exclude that for larger sizes one recovers self-averaging,
nevertheless our result is consistent with other studies which used
sample-averaged quantities~\cite{KPY,KM00,PY00,MP01} and also found
$\theta'\simeq 0$.  Independently of whether there is self-averaging
or not, we have provided evidence that the identity $G=1/3$ holds in
the thermodynamic limit in the whole spin-glass phase, a fact that
calls for a theoretical explanation in terms of the geometry and
energetics of the low-lying excitations. We have confirmed that $G$ and
$A$ can be used to locate the spin-glass transition, although in
models with time-reversal symmetry the usual sample-averaged
parameters provide a better determination. Finally,we have confirmed
the existence of a spin-glass phase transition at finite temperature
in three dimensions.

\begin{acknowledgments}
We acknowledge support from Spanish Ministerio
de Ciencia y Tecnolog\'{\i}a, Grants BFM2001-3525 (F.R.), AP98-36523875
(M.S.), Generalitat de Catalunya (F.R.). M.P. acknowledges support from
the European Commission, contract HPRN-CT-2002-00319 (STIPCO).
\end{acknowledgments}

\hspace{-2cm}

\end{document}